\documentclass{article}

\usepackage[nonatbib, final]{neurips_2019}
\usepackage[utf8]{inputenc} 
\usepackage[T1]{fontenc}    
\usepackage{hyperref}       
\usepackage{url}            
\usepackage{booktabs}       
\usepackage{amsfonts}       
\usepackage{nicefrac}       
\usepackage{microtype}      
\usepackage{listings}
\usepackage{graphicx}
\usepackage{amsmath}
\usepackage{color}
\usepackage{subfig}         
\usepackage{wrapfig, floatrow} 
\usepackage[font=small]{caption}
\usepackage{float}
\usepackage{afterpage}
\hypersetup{
  colorlinks = true, 
  urlcolor   = blue, 
  linkcolor  = blue, 
  citecolor  = black 
}

\newcommand{\figwithsidecaption}[2]{
\begin{figure}[H]
\floatbox[{\capbeside\thisfloatsetup{capbesideposition={right,top}, capbesidewidth=0.7\textwidth}}]{figure}[\FBwidth]
{\caption*{#1}}
{\includegraphics[width=2cm]{#2}}
\end{figure}
\vspace{-.5cm}
}

\title{Blue River Controls: A toolkit for Reinforcement Learning Control Systems on Hardware}

\author{
  Kirill Polzounov\thanks{Work done as an intern at Blue River Technology}\\
  University Of Calgary \\
  \texttt{kirill.polzounov@ucalgary.ca} \\
  \And
  Lee Redden \\
  Blue River Technology\\
  \texttt{lee.r@bluerivert.com} \\
  \And
  Ramitha Sundar \\
  Blue River Technology\\
  \texttt{ramitha.sundar@bluerivert.com} \\
}

\begin{document}
\maketitle

\begin{abstract}
Much of the recent success of Reinforcement learning (RL) has been made possible with training and testing tools like OpenAI Gym and Deepmind Control Suite \cite{OpenAIGym, DMSuite}. Unfortunately, tools for quickly testing and transferring high-frequency RL algorithms from simulation to real hardware environment remain mostly absent. We present \textit{Blue\ River\ Controls}, a tool that allows to train and test reinforcement learning algorithms on real-world hardware. It features a simple interface based on OpenAI Gym, that works directly on both simulation and hardware. We use \href{https://www.quanser.com/products/qube-servo-2/}{Quanser’s Qube Servo2-USB platform}, an underactuated rotary pendulum as an initial testing device. We also provide tools to simplify training RL algorithms on other hardware. Several baselines, from both classical controllers and pretrained RL agents are included to compare performance across tasks. \textit{Blue\ River\ Controls} is available at this https URL: 
\href{https://github.com/BlueRiverTech/quanser-openai-driver}{\texttt{github.com/BlueRiverTech/quanser-openai-driver}}
\end{abstract}

\section{Introduction}

Control systems are present in many facets of daily life ranging from coffee machines \cite{coffee} to cruise control in automobiles \cite{cruise-control} and have become a crucial part of the modern world.

Classical Control systems started becoming formalized in the early 1900s, mostly focusing on simple linear controllers like Proportional Integral Derivative (PID) control \cite{PID}. PID is a feedback control mechanism that assumes linear system dynamics \footnote{Although there are principled extensions to nonlinear systems such as feedback linearization and gain scheduling} and an action that is a linear combination of the state’s proportional, integral, and derivative components. Even with significant progress in the last few decades, most of the world is using linear control systems that were devised years ago. Although numbers vary, one study \cite{Honeywell-survey} estimates that 97\% of industrial controllers use PID control. Learned non-linear controllers have an opportunity to improve performance or allow to solve previously out of reach problems. This is especially true when applied to problems where traditional control is not well adapted like non-linear problems.

Reinforcement learning (RL). RL recently has made significant progress in solving complicated control problems. This is especially prominent in systems where the dynamics are too complicated to model and study. Much of this progress can be attributed to the tools simplifying the RL research process. These tools include auto-differentiation packages \cite{TF, Pytorch}, better simulators[Mujoco, Bullet], and RL testing toolkits \cite{OpenAIGym, DMSuite}. Further, RL shows promise in solving problems that can not be solved with traditional methods, including controlling a car directly from pixels \cite{raw-pixels-toycar}, or learn a latent embedding where the dynamics in the embedding is linear \cite{SOLAR}. Despite all the success, we are still missing good sample complexity, reliability, safety guarantees from RL \cite{recht-survey}.

Previous Work OpenAI Gym and Deepmind Control Suite are two popular software packages for training and testing RL agents. They provide a simple API that allows training reinforcement learning agent to interact in many different tasks (called environments). 

Reinforcement Learning Formalization Reinforcement learning is often formalized as a Markov Decision Process (MDP) or it’s generalization, a Partially Observed Markov Decision Process (POMDP). The MDP is characterized by state $s_t$, action $a_t$, reward function $r(s_t, a_t)$, the agent’s policy function $p_\theta(a_t|s_t)$ (parameterized by $\theta$), and a probability of transition $p(s_{t+1} |s_t , a_t )$. An RL agent’s goal is to maximize the expected reward $E_{(s,a) \sim p_\theta (s,a)} [r(t_0 ... t_T)]$

In the next section we describe the environments available in this work. In Section 3 we describe the extendable reinforcement learning interface that allows for the creation of new environment tasks. In Section 4, we document how to add additional hardware to work with Blue River Controls. Finally, section 5 presents future extensions to this work.

\section{Environments}
\subsection{Qube platform description}

\begin{figure}%
    \centering
    \subfloat[]{{\includegraphics[width=0.25\linewidth]{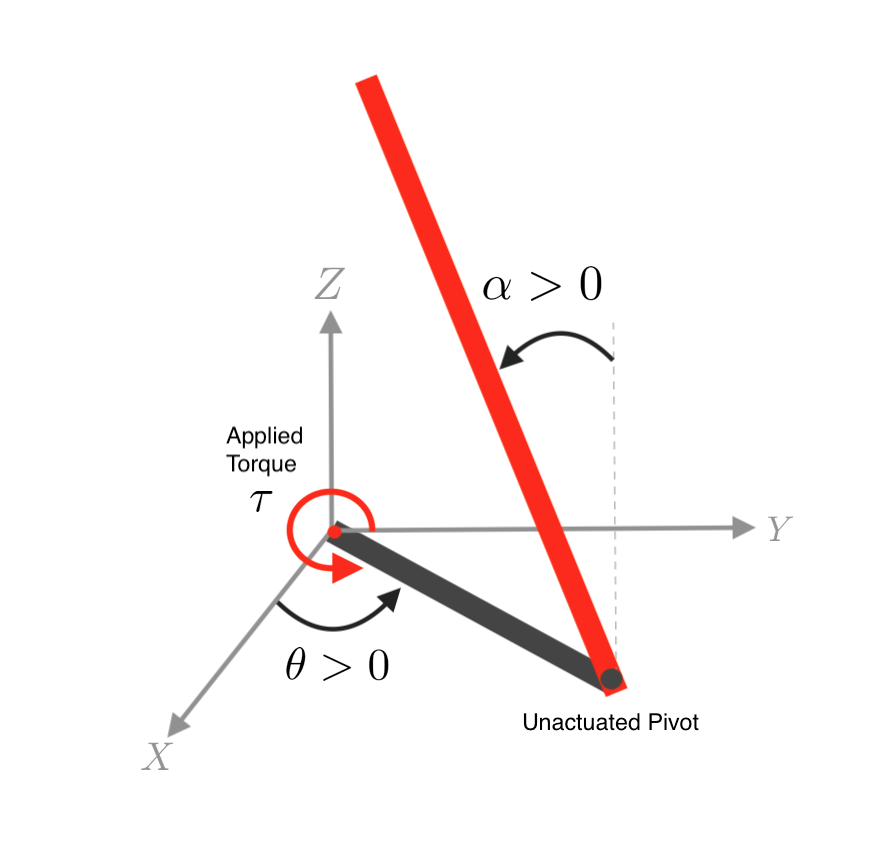}}}
    \qquad
    \subfloat[]{{\includegraphics[width=0.3\linewidth]{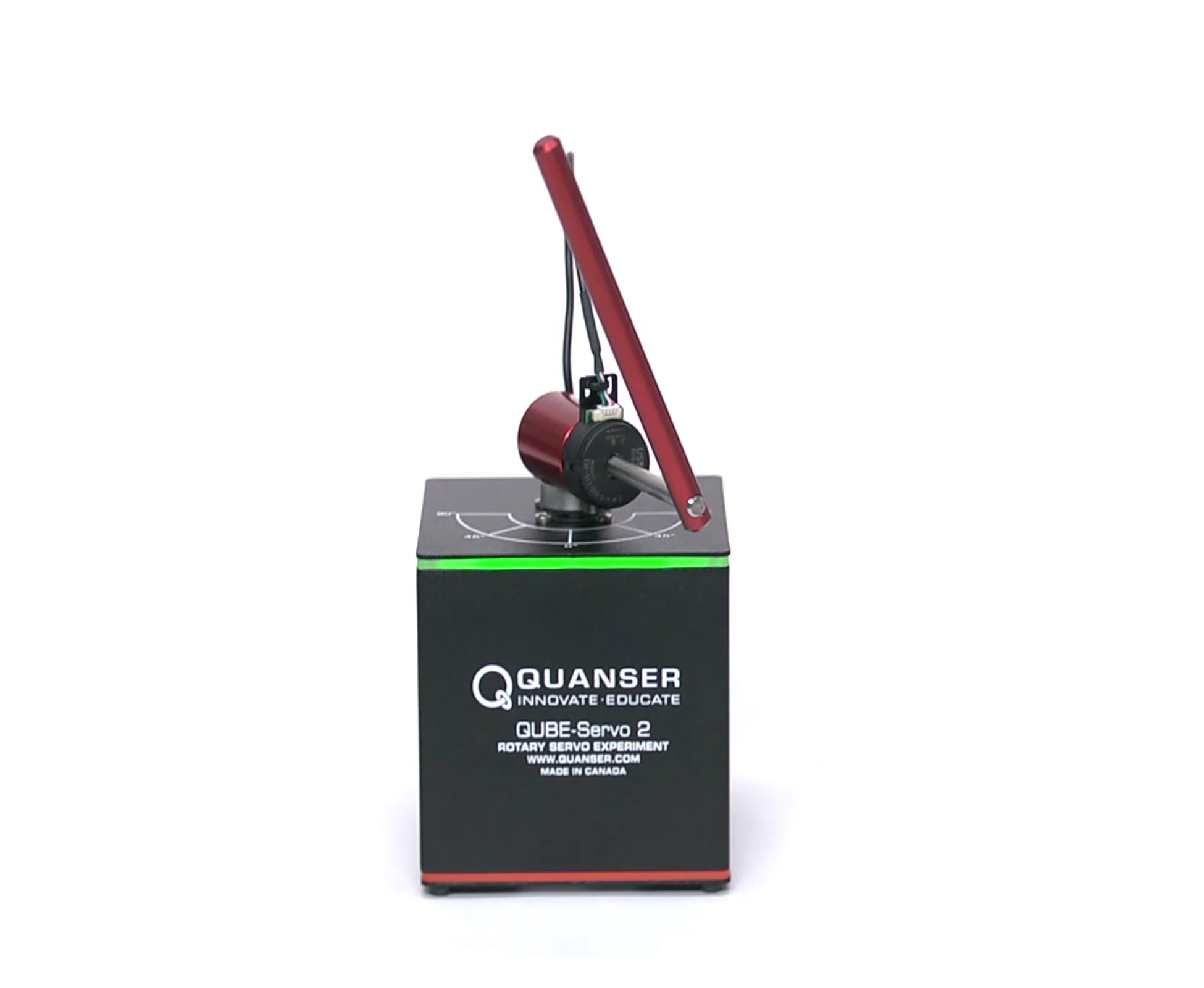}}}
    \caption{Qube Servo-2 USB}
    \label{fig:qube-servo2-usb}
\end{figure}

The Quanser Qube Servo2-USB (\textit{Qube}) is an underactuated rotary pendulum \cite{Rotary-pendulum-furuta} with a single DC motor \footnote{DC motor on hardware, and an ideal DC motor in simulation} controlling the rotary arm. A low friction link connects the pendulum to the rotary arm. The pendulum is unactuated but it can swing freely. All of the environments described in this work are based on the Qube, but are all easily extendable to new hardware as described in sections 3 and 4.

The \textit{Qube} uses the following conventions:

\begin{itemize}
\item
  $\alpha$: angle of \emph{pendulum} from upright.
\item
  $\theta$: angle of \emph{arm} from centered at the front.
\item
  $\dot{\alpha}$: angular velocity of \emph{pendulum} from centered
  at the front.
\item
  $\dot{\theta}$: angular velocity of \emph{arm} from centered at the
  front.
\end{itemize}

\subsection{RL Gym Tasks for the Qube platform}
In the Blue River Controls environments below, all tasks have an internal state space of 4 dimensions (the angles of the arm and pendulum as well as their velocities). The Follow tasks have 1 additional dimension representing $\theta_{target}$, the target rotary arm angle. All of the environments \footnote{Excluding the Qube Rotor} also have optional sparse rewards that make each task more difficult. 

The reward functions are designed to be easily comparable to each other. For example, Balance and Swingup tasks use the same reward function. In each episode the maximum reward is approximately $1$ \footnote{Reward is 1 for every step the pendulum is in the goal state, in some environments there is a minimum amount of time to reach the goal state, so it may be impossible to reach a reward of 1 for all timesteps.} for each timestep, and minimum reward of $0$ for the episode.

\figwithsidecaption{
   \textbf{Qube Dampen}: The pendulum begins at an arbitrary position. The goal is to apply torque to the rotary arm to quickly reach angle $\alpha=\pi$ and $\theta=0$ and remain stationary downwards.

  \centerline{
      $\displaystyle r = \frac{0.8 | \alpha + \pi | + 0.2 | \theta|}{\pi}$
  }
}{figures/dampen}

\figwithsidecaption{
    \textbf{Qube Balance}: Similar to the task given by \cite{Barto-1983}, with a rotary pendulum instead of a cartpole. The pendulum starts almost inverted with a small amount of randomness (in hardware this is innate, in simulation this is sampled from a Gaussian).
    
    \centerline{
      $\displaystyle r = \frac{0.8 | \alpha | + 0.2 | \theta |}{\pi}$
    }
}{figures/balance}

\figwithsidecaption{
    \textbf{Qube Swing-Up}: A task that requires an agent to swing up from a stationary downwards position and balance the pendulum afterwards.\\
    
    \centerline{
      $\displaystyle r = \frac{0.8 | \alpha | + 0.2 | \theta |}{\pi}$
    }
}{figures/swing}

\figwithsidecaption{
    \textbf{Qube Balance Follow}: An enhancement of the Balance task that adds a target angle for the rotary arm.
    
    \centerline{
      $\displaystyle r = max \Big(\frac{0.8 | \alpha | + 0.2 | \theta_{target} - \theta|}{\pi}, 0 \Big)$
    }
}{figures/balance-follow}

\figwithsidecaption{
    \textbf{Qube Swing-Up Follow}: An enhancement of the Swingup task that adds a target angle for the rotary arm.
    
    \centerline{
      $\displaystyle r = max \Big(\frac{0.8 | \alpha | + 0.2 | \theta_{target} - \theta|}{\pi}, 0 \Big)$
    }
}{figures/swing-follow}

\figwithsidecaption{
    \textbf{Qube Rotor}: The pendulum is initialized at rest (stationary downwards), the agent receives a reward for every $360^{\circ}$ rotation completed.
}{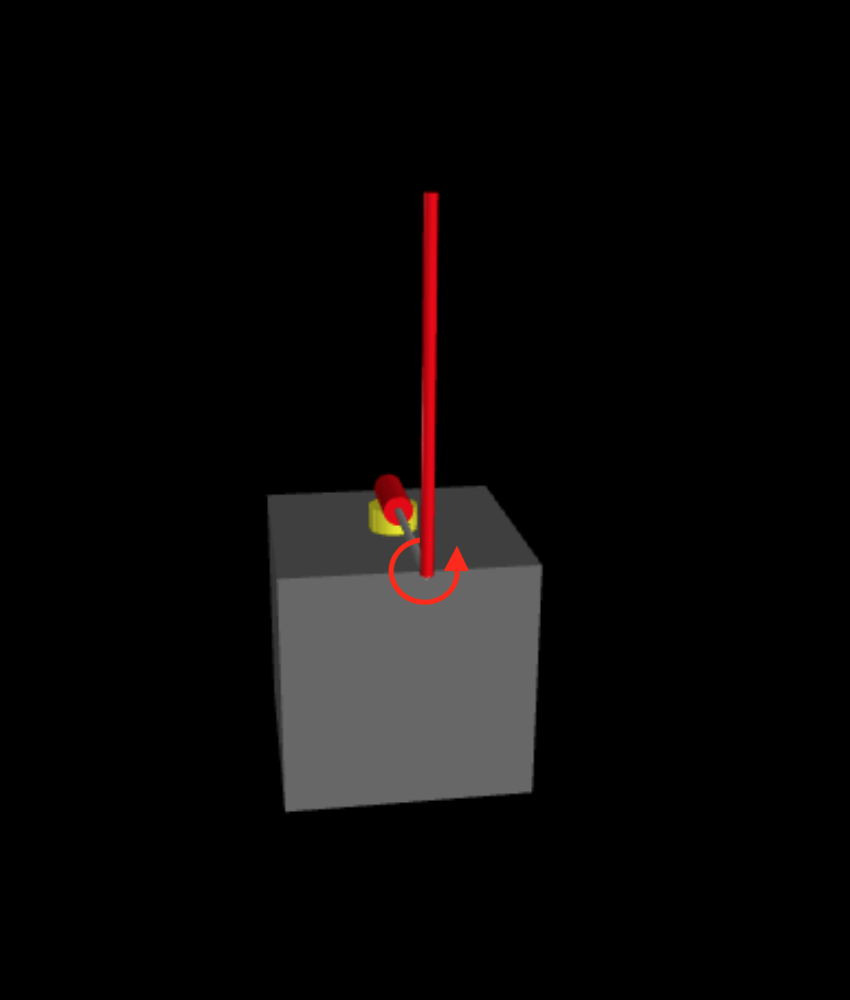}

\section{Reinforcement Learning Interface}

\begin{figure}
\centering
\includegraphics[width=10cm]{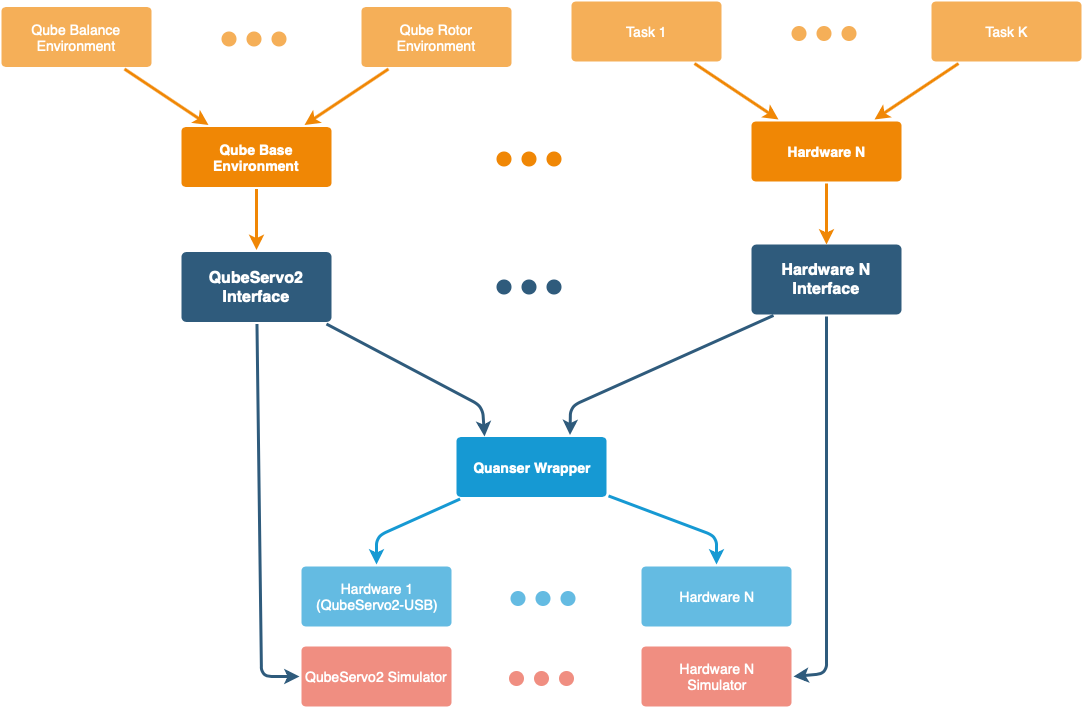}
\caption{High level overview of the architecture.}
\end{figure}

Figure 2: High level overview of the architecture.

Blue River Controls provides several OpenAI Gym environments that work both directly on hardware and a simple physics simulator. Blue River Controls interacts with hardware and simulators at a high level built on top of the OpenAI Gym API. This allows for several existing RL algorithms \cite{OpenAI-baselines, stable-baselines} to work out-of-the-box on the hardware.
          
\subsection{Qube Base Environment}
The Qube Base Environment contains the logic for directly running RL algorithms. The primary elements of this environment are the reset and step methods on the environment object (as well as render, and close). These methods implement the Partially observed markov decision process (POMDP) representing the task. The interface is identical for both simulation and hardware, with a simple option that can select how to run each task. \cite{DMSuite} separates the concept of hardware and goal by using the terms \textit{domain} and \textit{task} respectively. Domain defines all of the physics or hardware being used (in this work we only consider the Qube but have worked to make our library extendable to other hardware). Task refers to the goal that the agent must complete; this is described by our Qube environments in section 2.

\subsection{Reward Debugging }
Many reinforcement learning problems require a starting initial state or an initial state distribution. In simulation this is straightforward, either set the state in software or randomly sample from the initial state distribution. In hardware this is not possible. A controller must be used to reach an initial state.

Having a controller and a reinforcement learning agent quickly change between each other makes it difficult to understand to visualize which of the two is currently active. To simplify debugging RL algorithms Blue River Controls provides an LED light on hardware that displays the current state of the environment. Yellow represents a reset controller. Green represents a state that corresponds to high reward, and red represents a state corresponding to low reward.

\section{Hardware-Python Interface}

\subsection{Quanser Wrapper}

We provide a simple hardware wrapper around the Quanser’s hardware-in-the-loop software development kit (HIL SDK) to allow for easy development of new Quanser hardware. To connect to the hardware we use a module written in \href{https://cython.org}{{Cython}}. The internal \texttt{QuanserWrapper} class handles most of the difficult aspects of interacting with hardware, including the timing (using a hardware timer), and ensuring the data sent to hardware is safe and correct, where safety corresponds to safe operating voltage and current for the specified hardware.

\newpage
A simple example of how to use the Qube with python.

\begin{lstlisting}[language=Python]
with QubeServo2(frequency=250) as qube:
    currents, encoders, others = qube.action(np.random.randn())
    print(encoders)
\end{lstlisting}
\begin{lstlisting}[language=Bash]
>>> ([0.0, 0.0])
\end{lstlisting}

The \texttt{QuanserWrapper} class allows for easy extendability by defining the communication channels being used on hardware and defining the name. These can be found in the documentation given by Quanser for the specific hardware.

\subsection{Quanser Interface}
For all tasks to work on both hardware and simulation equally well, we define an interface that does some of the work required to match a simulator to a piece of hardware. In simulation, it is trivial to set the state to an arbitrary value. Running RL training on hardware makes this significantly more complicated. Simple environments like the Balance task must be set into an initial state by running a separate (non-reinforcement learning) controller that can swing up the pendulum first, then allow the RL agent to train. The Quanser Interface implements all of the logic required for a reset to an initial state, to retrieve the full state (the Qube only has sensors for the angles, not the angular velocities), and any other required methods to match hardware and simulation.

\section{Benchmarks}
We provide two forms of benchmarks in Blue River Controls. We provide models, checkpoints, and training code using stable baselines \cite{stable-baselines} for Proximal Policy Optimization (PPO) \cite{ppo}. PPO is an on-policy model-free policy gradient algorithm. It is a commonly used algorithm and is competitive against many other model-free algorithms due to low variance and consistent evaluation runs. In addition to RL baselines we also provide classical control baselines for various tasks. While comparing to state-of-the-art RL baselines is useful, to encourage adoption in real world systems, reinforcement learning must show benefits that outweigh potential costs (such as training and difficulty of implementation) when compared to traditional control systems. Benchmarks are provided on Github here: \href{https://github.com/kirill5pol/qube-baselines}{Qube Baselines}.

\section{Future Work}

\subparagraph{More devices}
Due to the extendability of the work, it would be very simple to add new
quanser hardware, this ranges from \href{https://www.quanser.com/products/quanser-aero/}{{UAV testing tools}},
to \href{https://www.quanser.com/products/active-suspension/}{{active suspension}} or \href{https://www.quanser.com/products/3-dof-crane/\#overview}{{cranes}}.

\subparagraph{Optimal control algorithms}
Many modern control systems outperform reinforcement learning in many cases, unfortunately, most research in RL only compares the performance between RL algorithms rather than more broadly. It would be beneficial for the community to have more broad baselines for comparison. Iterative linear quadratic regulator (iLQR) and model predictive control (MPC) \cite{mpc} \footnote{A downside of MPC compared to (policy-function based) RL is MPC must replan every timestep, which is becomes challenging in real time inference (such as 250 Hz or above).} are two such algorithms that have shown much success on a variety of tasks. These would be very interesting to compare against the current state of the art in RL.

\subparagraph{More accurate simulators and different accuracy options}

\begin{itemize}
\item
    Contact-based physics simulators like Mujoco \cite{mujoco} and Bullet \cite{bullet} add more realism than the physics simulator we currently use.
\item
    Domain randomization has been shown to be a viable method for transfer from simulation to real hardware \cite{Domain-Randomization, Learning-Dexterity}.
\item
    Allowing for simulator accuracy settings that can be turned on or off with a flag on a simple problem. Some possible options include: discretization that match the sensors, hardware delay, and more realistic actuator output, then see how each option affects transfer performance. By starting on a simpler environment This would allow for insights into which portions of simulators are crucial to transfer onto hardware.
\item
    By studying a relatively simple problem like the Qube, it may be easier to observe the effects of different options for accuracy and which portions of simulators to add noise (in the case of domain randomization).
\end{itemize}

\bibliographystyle{unsrt}
\bibliography{main} 

\end{document}